\documentclass[twocolumn,english]{revtex4}
\usepackage[T1]{fontenc}
\usepackage[latin9]{inputenc}
\usepackage{amstext}
\usepackage{amsmath}
\usepackage{graphicx}
\usepackage{esint}
\usepackage{times}   

\makeatletter

\@ifundefined{textcolor}{}
{%
 \definecolor{BLACK}{gray}{0}
 \definecolor{WHITE}{gray}{1}
 \definecolor{RED}{rgb}{1,0,0}
 \definecolor{GREEN}{rgb}{0,1,0}
 \definecolor{BLUE}{rgb}{0,0,1}
 \definecolor{CYAN}{cmyk}{1,0,0,0}
 \definecolor{MAGENTA}{cmyk}{0,1,0,0}
 \definecolor{YELLOW}{cmyk}{0,0,1,0}
 }

\makeatother

\usepackage{babel}
\begin{document}

\title{Stochastically Projecting Tensor Networks}

\author{Bryan K. Clark}
\affiliation{Department of Physics, University of Illinois at Urbana-Champaign, 
1110 West Green St, Urbana IL 61801, USA}
\author{Hitesh J. Changlani}
\affiliation{Department of Physics, University of Illinois at Urbana-Champaign, 
1110 West Green St, Urbana IL 61801, USA}
\date{\today}

\begin{abstract} 
We apply a series of projection techniques on top of tensor networks
to compute energies of ground state wave functions with higher accuracy 
than tensor networks alone with minimal additional cost.
 We consider both matrix product states as well as tree 
tensor networks in this work.  Building on top of these approaches, we apply 
fixed-node quantum Monte Carlo, Lanczos steps, and exact projection. 
We demonstrate these improvements for the triangular lattice Heisenberg model, 
where we capture up to  $57\%$ of the remaining energy not captured by 
the tensor network alone. We conclude by discussing 
further ways to improve our approach.
\end{abstract}

\maketitle
\section{Introduction}
Numerical methods are important tools for understanding strongly 
correlated systems. One such approach is the density
matrix renormalization group (DMRG)~\cite{dmrg_white,White1993} which is 
particularly powerful for one-dimensional quantum systems. 
In recent years, DMRG has also had many successes in 
simulating quasi one-dimensional ladders which is a way of approaching 
the two-dimensional limit~\cite{Stoudenmire_White_2D_DMRG}. 
Unfortunately, the bond-dimension $D$ required in DMRG (the parameter 
that determines the accuracy), scales exponentially with the width of the ladders. 
The underlying variational wavefunction that is optimized by the DMRG algorithm is a 
matrix product state (MPS)~\cite{Ostlund-Rommer}. 
Higher dimensional generalizations of the MPS idea 
have led to development of new tensor network (or related) 
approaches~\cite{TPS_review} 
such as TTN~\cite{Shi_Vidal,Evenbly_Vidal_tree,Murg_TTN}, MERA~\cite{Vidal_MERA}, 
CPS~\cite{Huse_Elser,Changlani_CPS,mezzacapo,Marti} and PEPS~\cite{tps_nishino,Nishino_3D_classical}. 
These are known to be better than 
DMRG at capturing the entanglement structure of physical systems and 
hence can produce better energies for the same bond-dimensions. 
Unfortunately, the computational expense of optimizing these wavefunctions 
and calculating observables (such as the energy) scales formidably with $D$. 
This motivates us to look for approaches which take a low bond-dimension ansatz
generated by DMRG or other tensor network methods and improve upon them. 

One such approach is to apply particular projectors $\hat{P}$ to a 
tensor network wavefunction $\Psi$ which generates a new wavefunction $\hat{P}\Psi$ which 
is closer to the ground state. In fact, this technique is often used to optimize a matrix product
state through evolution in imaginary time~\cite{tebd_Vidal}. 
In this paper, we will explore the ways in which quantum Monte Carlo methods 
can help generate or evaluate projectors $\hat{P}$ producing a (possibly stochastic) 
representation of $\hat{P}\Psi$. 

There has been fruitful progress when
Monte Carlo and tensor network approaches have been combined. Examples include Refs.~\cite{Wang_TNQMC,Ferris_Vidal,Sandvik_Vidal_MPSQMC} 
where Monte Carlo is used to evaluate observables or optimize tensor networks 
and Ref.~\cite{Pollmann_MPS_Slater} where a MPS is used as one component of a larger 
variational ansatz. Additionally, for the square $J_1-J_2$ model, 
Ref.~\cite{de2000incorporation} used an approach similar to 
one of the directions explored here. 

In this paper, we explore the application of three projectors $\hat{P}$ to tensor networks:
a stochastic application of 
(1) $\hat{P}_\textrm{exact}=  \exp[-\beta H]$ absorbing the sign problem for small imaginary time 
$\beta$, 
(2) $\hat{P}_\textrm{Lanczos} \equiv 1+\alpha H + \Delta H^2$ 
~\cite{Heeb_Rice,Sorella_Lanczos_step} (as well as generalizations thereof)
and 
(3) $\hat{P}_{FN}$, a sign problem free projector 
proposed and studied in Refs.~\cite{Bemmel1994,Haaf1995}. 
We also discuss the possibility of chaining these different projectors. 

To demonstrate these approaches, we consider the spin 1/2 Heisenberg Hamiltonian
\begin{equation}
H=\sum_{\langle i,j\rangle}S_{i}\cdot S_{j}
\end{equation} 
on the two-dimensional triangular lattice
where $S_i$ refers to the spin 1/2 operator on site $i$ and 
$\langle i,j \rangle$ are nearest neighbors. 
This system is frustrated and has a sign problem in the Ising basis. 
The triangular lattice has been commonly used as a benchmark for testing new tensor network 
algorithms~\cite{parallel_DMRG,White_triangular} and is considered challenging 
for these methods owing to the high coordination number of every site.
We use cylindrical boundary conditions with an equal number of unit cells 
in each direction.

In the remainder of the paper, we will explain our methods and demonstrate that these projectors 
typically recover between 15 and 57 percent of the energy missed 
by the tensor networks themselves.

\section{Tensor Network States}
Consider the full many body wavefunction $|\Psi \rangle$ 
expanded in a basis of configurations $|c\rangle$ 
with coefficients $\Psi[c]$
\begin{equation}
	|\Psi \rangle = \sum_{c} \Psi[c] |c \rangle
\end{equation}
For a $N$ spin system the chosen configurations could be Ising states $|c_1 c_2...c_N \rangle$
with $c_i$ referring to the spins on site $i$. 
Then a tensor network is a variational ansatz for $\Psi[c]$, which can be generically written as, 
\begin{equation}
\Psi[c]=\prod_{i}M_{\vec{\alpha}}[i;c_{i}] 
\end{equation} 
where each
$M_{\vec{\alpha}}[i;c_{i}]$ is indexed by the site $i$ and the value
of the physical spin $c_{i}$. 
The values of $\vec{\alpha}$ are arbitrary tensor indices $\vec{\alpha}=\{\alpha_{0}...\alpha_{n}\}$. 
In this notation, the product implies a contraction over all shared tensor indices between tensors.
In any tensor network scheme, the indices are chosen so that contraction
over them leaves $\Psi[c]$ a scalar. Each tensor index has a maximal
dimension of size $D$ which is a tuning parameter that extrapolates
between a product state ($D=1$) and the exact wave-function $(D\rightarrow\infty$).
Smaller $D$ corresponds to lower entanglement. Different forms of
tensor networks differ in the way in which the tensors are connected.

\subsection{Matrix Product States (MPS) and Tree Tensor Network (TTN) States}
In this work, we will focus on tensor networks whose indices
are connected into paths (MPS) as well as those that are connected
into trees (TTN). While MPS has been used extensively for a wide variety of 1D and quasi 2D
systems~\cite{Schollwock}, the use of TTN is relatively recent.
Several authors have employed them to study tree geometries~\cite{Changlani_tree_DMRG,
Changlani_percolation,Depenbrock_TTN,Li_TTN}, 2D geometries 
and quantum chemistry~\cite{Nakatani_Chan_TTN,Murg_TTN}.
It is to be noted that the flavor of TTN used here (and in all the above mentioned works) 
is such that the physical spins occupy the vertices of a tree.~\footnote{The flavor of TTN used here 
is quite distinct from the TTN introduced by Vidal~\cite{Evenbly_Vidal_tree}, which has 
all the physical sites as the leaves of a tree and a hierarchical arrangement of additional 
tensors that introduce entanglement. This is basically a MERA network~\cite{Vidal_MERA} without 
the disentangling operations.}

\begin{figure}
\includegraphics[width=\linewidth]{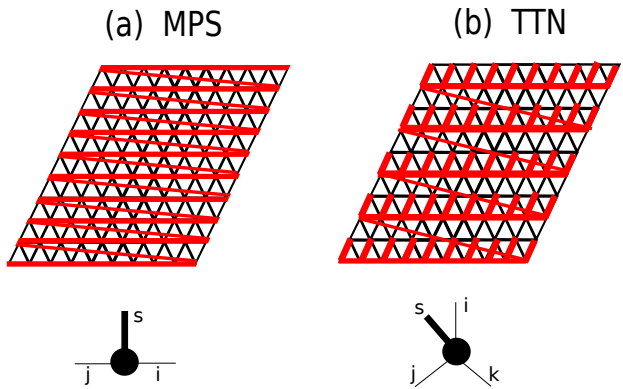}
\caption{(Color online) Tensor networks shown in red used in this work 
(a) Matrix Product State which is optimized with the DMRG algorithm. 
This involves mapping the sites on a 2D lattice onto a "snake" used to define the sweep order. 
Each site has two matrices associated with it (one for each spin value) with auxiliary 
indices $i$ and $j$ which are summed over. 
(b) A particular realization of a tree tensor network with coordination 3 and 
coordination 1 vertices, which is a particular map of 
spins on a 2D lattice onto a tree (loopless) network. 
Each coordination 3 site has three indices $i,j,k$ which are summed over. 
The figure also shows the underlying triangular lattice (in black) 
on which the nearest neighbor Heisenberg Hamiltonian is defined. 
Periodic boundaries along the x-direction have not been shown.}
\label{fig:mps_ttn}
\end{figure}

Figure~\ref{fig:mps_ttn} shows the networks and the tensors we use. 
The trees we have selected have coordination 3 and 4. The MPS wavefunctions are 
generated using the ITensor library~\cite{ITensor} and the 
TTN wavefunctions have been generated from our Tree-DMRG code, 
(details of which have already been discussed elsewhere~\cite{Changlani_tree_DMRG}). 

It is important to note that the "optimal mapping" from a tree to a 2D system is not a 
trivial problem, since there is a subtle interplay between short and long range entanglement 
necessary for getting gains in the energy. Practitioners of TTN have typically 
used heuristic algorithms to arrive at a reasonable mapping. Here we select a network that 
covers some short range motifs not efficiently covered by the MPS. 
These networks can be contracted computationally efficiently using $O(D^z)$ operations 
with $z$ being the maximum degree
of the tree. 

As an example, for $D=8$, we find that switching from MPS to TTN can gain 
$30\%$ (for coordination 3) and $40\%$ (for coordination 4) of the missing energy.  

While we focus on these particular tensor networks in this work,
our conclusions likely generalize to other tensor networks such as PEPS, 
where contraction is even more computationally costly (exponential in $D$ 
if performed exactly and $D^6$ if performed approximately~\cite{Wang_TN_MC}) and where it 
is even more important to improve on lower bond dimensions. 
Since in the limit of large $D$, any state can be represented by a tensor network $\Psi$, 
the state $\hat{P} \Psi$ is also guaranteed to converge 
to the correct energy in this limit. 

\subsection{Use of quantum numbers and gains from post projection}
Since DMRG is a $D^3$ algorithm, several efforts are made in its implementation 
to reduce the prefactor associated with the calculation.  
One such improvement, at the cost of some extra book-keeping, 
is the use of the good quantum numbers such as total $S_z$ and/or total 
particle number.

\begin{figure}[htpb]
\includegraphics[width=0.96\linewidth]{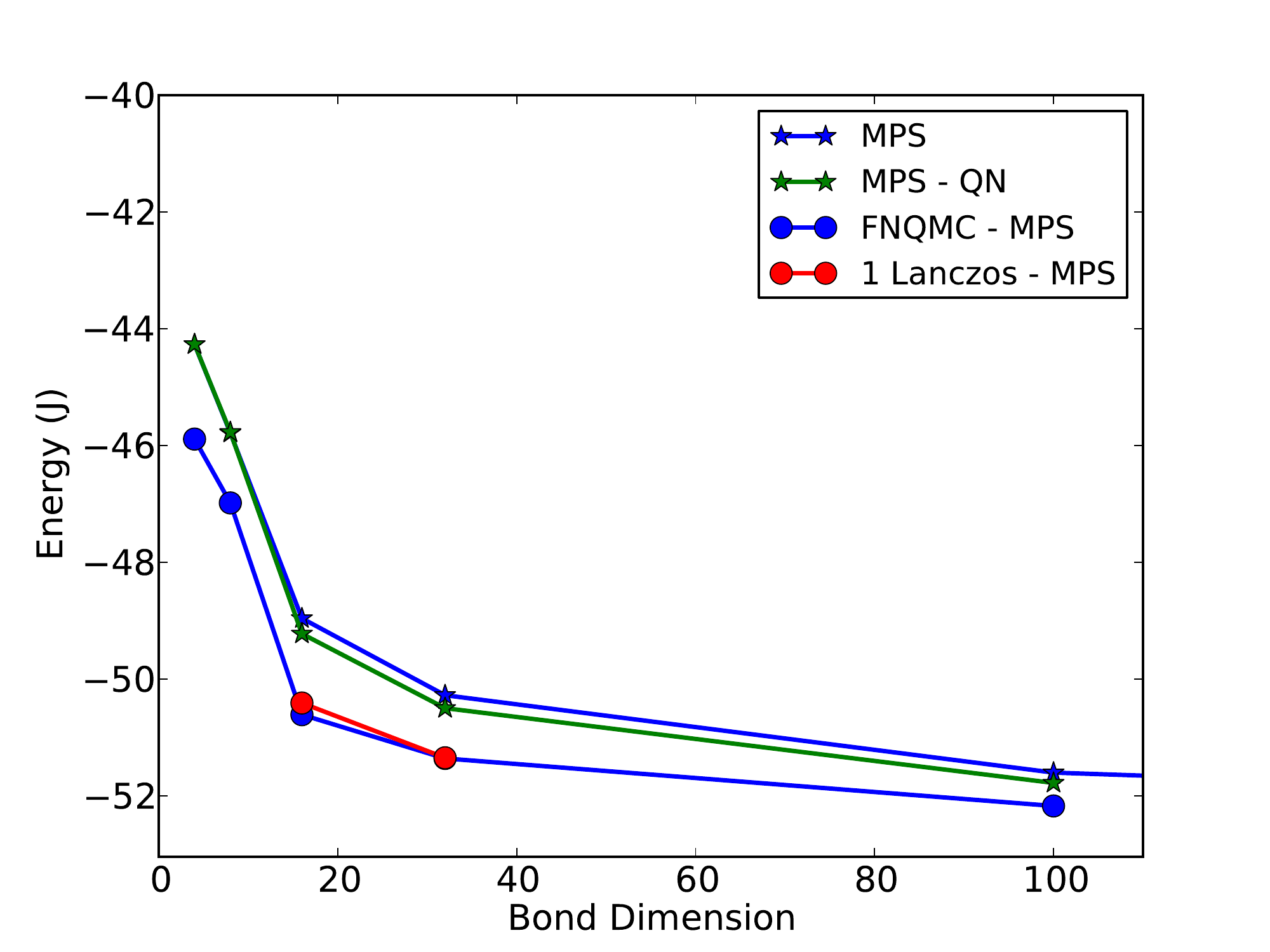}
\caption{Total ground state energy obtained from MPS of various bond dimensions generated 
without the use of total $S_z$ quantum number. 
The projection onto total $S_z = 0$ sector is reinstated in the QMC calculation 
by calculating the energy over samples with that symmetry. The legend shows different projection
methods and their respective decrease in energy as a function of bond-dimension.}
\label{fig:noQN}
\end{figure}

However, it is not apriori obvious that this is 
necessarily advantageous, because respecting a global symmetry may introduce
additional entanglement in the state. Instead, breaking symmetries and then 
restoring them may allow for lowering of the energy 
for a given bond dimension. This has recently been observed in the context of Hartree Fock wavefunctions~\cite{Scuseria_wavefunction}. 
One can reinstate the symmetry using Quantum Monte Carlo (see Figure~\ref{fig:noQN} for the gain in energy before and 
after projection), by sampling configurations which have definite $S_z$ and/or particle number. 
We see evidence for this assertion, when we compare 
MPS with and without the use of any good quantum numbers (see Figure~\ref{fig:noQN} and \ref{fig:QN} for a comparison). 
Also note that such a global post projection of the tensor network is not necessarily very straightforward 
in the MPS framework alone, but easy to perform in QMC.

\section{Projector Quantum Monte Carlo}
QMC methods work by filtering out excited states from a given trial state $\Psi_T$ 
by applying to it, a projector,
\begin{equation}
	\hat{P} | \Psi_T \rangle = \lim_{\beta \rightarrow \infty} \exp(-\beta H) |\Psi_T \rangle  \propto | \Psi_0 \rangle 
\end{equation}
where $| \Psi_0 \rangle$ is the ground state. 
QMC methods are most efficient in systems where there is no "sign problem"; 
for example, when all the off-diagonal matrix elements of $H$ are negative 
in a chosen basis. Otherwise, the stochasticity in the method 
causes the variance to grow exponentially with system size 
and inverse temperature $\beta$ arising from a cancellation of large positive and negative terms. 
This makes it difficult to reliably measure observables for system sizes of interest.
However, as we will see later in this section, some approximate "sign-problem free" projectors 
can also be used, at the expense of a variational bias in the final answer.

To quantify the gains from a generic projector methods using a MPS wavefunction 
(henceforth abbreviated as QMC+MPS), we define the fraction of missing energy as 
\begin{equation}
f=\frac{E_{MPS}[D]-E_{QMC+MPS}[D]}{E_{MPS}[D]-E_{MPS}[\infty]}
\label{eq:fraction_recovered}
\end{equation}
where $ E_{MPS}[D]$ is the energy of the MPS of bond dimension
$D$, $E_{QMC+MPS}[D]$ is the energy of the QMC using the same
MPS of bond dimension $D$ and $E_{MPS}[\infty]$ is taken to be the
energy 
extrapolated
to $D\rightarrow\infty$, which is found to be -53.04 $J$. 

\subsection{Exact Projection}
Projector Monte Carlo methods work by producing a stochastic 
representation of the entire many body wavefunction. 
The important sampled wavefunction ($\Psi \Psi_T $) 
is represented by walkers, entities that carry weights $w_c$ and signs $s_c$ 
associated with configurations (or basis states, denoted by $c$) in
the Hilbert space. 

The projected (or mixed) energy can be written as 
\begin{equation}
	E = \frac{ \langle \Psi | H | \Psi_T \rangle} {\langle \Psi | \Psi_T \rangle }
\end{equation}
which can be recast as,
\begin{equation}
	E = \frac{ \sum_{c} \left| \Psi(c) \Psi_T(c) \right| \text{sgn}(\Psi(c)\Psi_T(c)) E_L[c]} {\sum_{c} \left| \Psi(c) \Psi_T(c) \right| \text{sgn} (\Psi(c)\Psi_T(c))}
\end{equation}
where $\left| \Psi[c] \Psi_T[c] \right|$ is the probability distribution being sampled by 
the projection process and $E_L[c] \equiv \langle c \left|  H \right| \Psi_T\rangle / \langle c | \Psi_T\rangle $
which is referred to as the "local energy". 
Thus, we compute the estimator 
\begin{equation}
E = \frac{ \langle s_i E_L[i] \rangle} { \langle s_i \rangle}
\end{equation} 
where $i$ are the samples accumulated during the projection process. Since this estimator corresponds to 
the energy of the wavefunction $\exp(-\beta H/2) | \Psi_T \rangle$, 
it is guaranteed to be a variational upper bound of the true ground state.   Notice that each calculation
of the form $E_L[i]$ takes $O(N)$ contractions of $\Psi_T$ where $N$  is the number of sites.

\begin{figure*}[htpb]
\includegraphics[width=\linewidth]{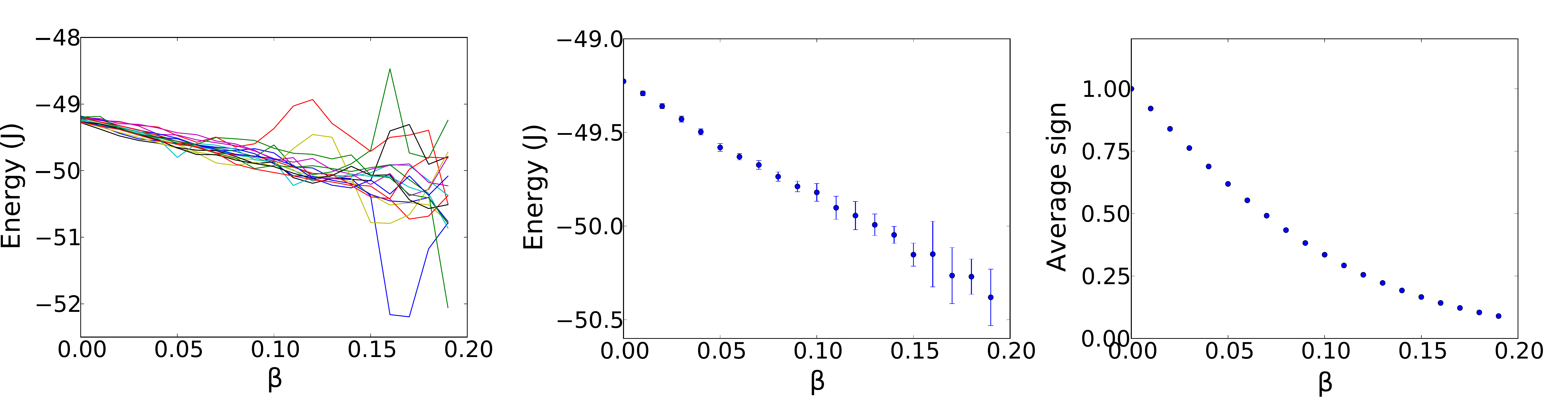}
\caption{Left: Energy of $\exp[-\beta H/2] |\Psi_T\rangle$ as a function of $\beta$ computed
using projector quantum Monte Carlo for 16 different runs of 100,000 walkers each. Center: The energy vs $\beta$ 
after averaging the runs. Right: Average sign $\langle s_i \rangle$ as a function of $\beta$ computed using
projector quantum Monte Carlo. The average sign decays exponentially. The trial wavefunction used in the initial VMC 
is the $D=16$ MPS with no quantum numbers.  
}
\label{fig:release}
\end{figure*}

We apply the exact projector (for small $\beta$) using (release-node) QMC ~\cite{Ceperley_release}
testing it on the 10$\times$10 triangular lattice.  
This is accomplished by first sampling, via variational Monte Carlo, 
the distribution of walkers on configurations $c$ from $|\langle \Psi_T | c \rangle|^2$. 
Our $\Psi_T$ is the MPS wavefunction with $D=16$ 
(without quantum numbers).  The results are shown in Figure~\ref{fig:release}. The starting energy 
is  the variational energy of this state, which is calculated to be around -49.2 $J$.

Once the exact projection is started, the energy is found to decrease but the statistical error quickly increases. 
We find that beyond $\beta \approx 0.2 $, the errorbars become unacceptably large. 
These runs have been calculated with $1.6 \times 10^6$ walkers. 
Error bars can be reduced with more cores (statistics) and efficient algorithms~\cite{Alavi_FCIQMC,SQMC,Clark_partial_node}, 
but we do not expect to go to significantly larger $\beta$ because the average sign 
decays to zero exponentially (as seen in Figure~\ref{fig:release}).  

\begin{figure}[htpb]
\includegraphics[width=\linewidth]{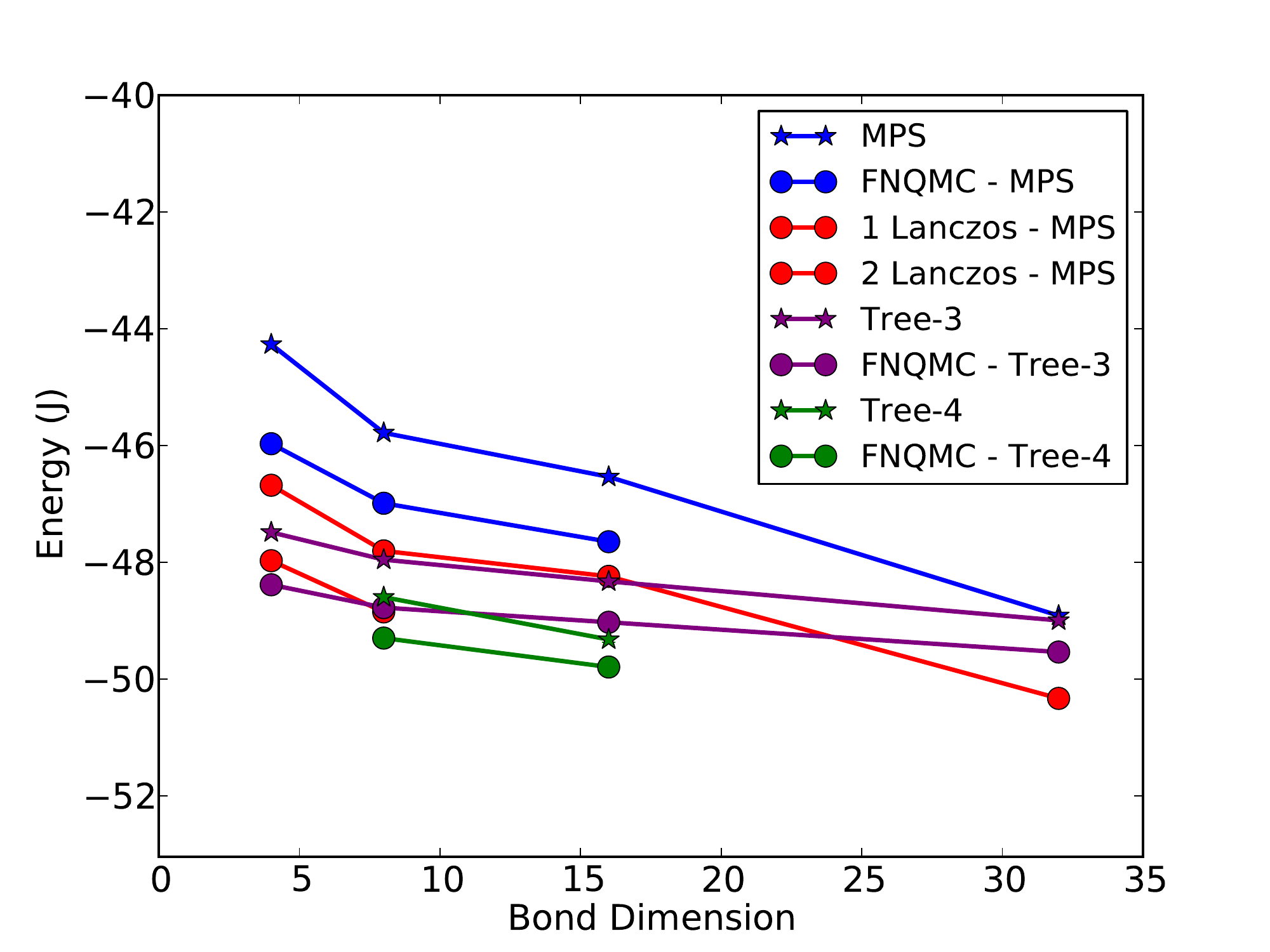}
\caption{Total ground state energy obtained from MPS and TTN of various bond dimensions generated 
with the use of total $S_z$ quantum number. Legend shows different projection
methods and their respective decrease in energy as a function of bond-dimension.}
\label{fig:QN}
\end{figure}

\subsection{Fixed Node}
Because exact QMC projection scales exponentially, 
applying $\hat{P}_\textrm{exact}$ eventually becomes untenable. 
Instead, one can apply fixed node QMC (FNQMC) which takes a Hamiltonian $H$ 
with a sign problem as well as a guess wavefunction 
$\Psi_{T}$ for the ground state of $H$ and generates a new 
(importance-sampled) effective Hamiltonian $H_{FN}$ 
that has no sign problem~\cite{Bemmel1994,Haaf1995}. 
This new Hamiltonian has off diagonal elements,
\begin{subequations}
\begin{eqnarray}
\left(H_{FN}\right)_{ij} & = & H_{ij}\frac{\psi_T(c_j)}{\psi_T(c_i)} \;\;\text{ if $ H_{ij}\frac{\psi_T(c_i)}{\psi_T(c_j)} <0 $}\\
                         & = & 0 \;\;\;\;\;\;\;\;\;\;\;\;\;\;\;\;\;\;\; \text{ otherwise      }
\end{eqnarray}
\end{subequations}
and whose diagonal elements are modified as,
\begin{equation}
\left(H_{FN}\right)_{ii}= H_{ii} + \sum_{\text{j $\in$ s.v.}} H_{ij} \frac{\psi_T(c_j)}{\psi_T(c_i)}
\end{equation}
where "s.v." refers to only the sign violating terms (terms that have $ H_{ij}\frac{\psi_T(c_i)}{\psi_T(c_j)}>0$). 

The ground state $\Psi_{FN}$ of $H_{FN}$ is an approximation to the ground state 
$\Psi_{0}$ of $H$ which becomes exact in the limit where $\Psi_{T}=\Psi_{0}$. 
In addition, the energy is variationally upper-bounded so that 
$\langle\Psi_{FN}|H_{FN}|\Psi_{FN}\rangle\geq\langle\Psi_{FN}|H|\Psi_{FN}\rangle\geq\langle\Psi_{0}|H|\Psi_{0}\rangle$. 
Therefore an effective use of FNQMC requires generating a good guess $\Psi_{T}$. 
The only requirement for $\Psi_{T}$ is the ability 
to quickly evaluate the amplitude $\Psi_{T}[c]$ for arbitrary configurations $c$. 
The general lore is that the sign structure of $\Psi_{T}$ 
is the most important aspect of the guess with a secondary importance
on the amplitudes (in the continuum, it can be shown that only the
sign structure matters). We use tensor networks, then, for $\Psi_{T}$. 

As shown in Figure \ref{fig:QN}, we apply FN-QMC on MPS and TTN 
of fixed quantum numbers ($S_z=0$) with various bond dimensions. 
Calculating $\langle\Psi_{FN}|H_{FN}|\Psi_{FN}\rangle$, we find that the 
tensor network ansatz for low bond dimensions lowers the energy by about 
1$J$. The tree ansatz provides a better starting point 
for low $D$ and improves upon increasing coordination.
These ansatz are able to capture about 15-20\% of the remaining missing energy.

We find that the quantum number free MPS (followed by QMC post projection), 
provides a much better starting point than its quantum number counterpart,
and with FNQMC captures a much larger fraction of the energy, about $40 \%$.

Note that the fixed node energies underestimate the quality of the wavefunction 
as the true energy, $\langle\Psi_{FN}|H|\Psi_{FN}\rangle$ is
actually even closer to the ground state energy. 

%
\subsection{Lanczos Step}
The final projector we consider is applying $\hat{P}_\textrm{Lanczos}$.
In order to apply this, we need to compute the optimal values of 
$\alpha$, $\Delta$ which we take to be those that minimize the variational energy.
Consider the energy 
\begin{equation}
\langle E \rangle = \frac{\langle \Psi (1 + \alpha H + \Delta H^2)  | H | (1 + \alpha H +\Delta H^2) \Psi \rangle}{\langle \Psi (1 + \alpha H + \Delta H^2) | (1 + \alpha H + \Delta H^2) \Psi \rangle}
\end{equation}
This requires computing terms of the form $\langle \Psi  |H^n | \Psi \rangle$ for $n \leq 5$. 
Computing these terms can be done either in the context of a matrix-product state/matrix-product operator 
formalism or within quantum Monte Carlo. In the MPS/MPO formalism, 
one can generate a MPO operator for $H$ which consists of four matrices $M^{\sigma_i,\sigma_j}$ 
for each site, each of bond dimension $\chi$ (for the $10\times 10$ lattice, $\chi=35$). 
One can then apply the MPO to the MPS computing a new MPS $H^{n/2}|\Psi \rangle$ by contracting over the 
physical degrees of freedom. This new MPS will have a bond dimension of $\chi^n/2 D$.  Then to compute 
$\langle \Psi H^n \Psi \rangle$ we simply compute the overlap of $| H^{n/2} \Psi \rangle$ with itself 
(appropriate modifications can be made for odd $n$).  Using this viewpoint, the computational complexity of this procedure
is then $(D^3 \chi^{3n/2})$ but the constant prefactor for this approach is extremely good because it
requires only a single overlap calculation. Further improvements from the sparsity of the MPO can potentially speed  
this up.
Notice that this is significantly cheaper than 
doing a DMRG calculation at that bond-dimension.

\begin{figure*}
\includegraphics[width=\linewidth]{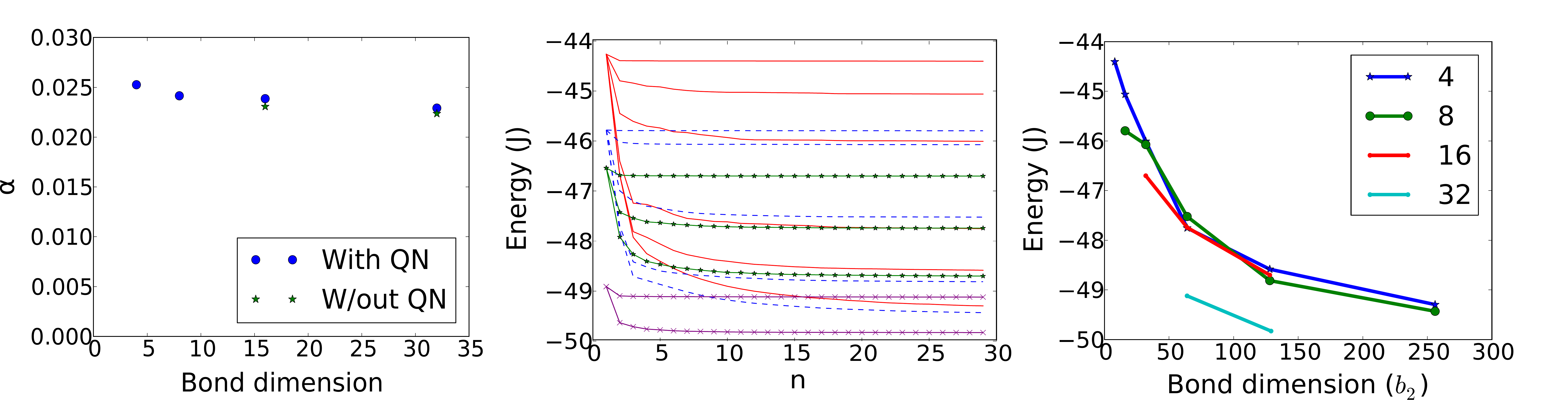}
\caption{Left: Optimal exact 1-Lanczos wavefunction parameter $\alpha$ obtained by minimizing the variational energy, 
as a function of the bond dimension $b_1$ of the MPS. 
Center: Energy as a function of basis elements $n$ in the approximate Lanczos method. 
The various colors correspond to choices of $b_1$= \{ 8 (solid red), 16 (dotted blue) 32 (starred green), 64 (purple x's).\}
For each color/symbol the lines from top to bottom represent choice of $b_2$ starting from $2 b_1$
and incrementing in multiples of 2. Right: Energy for $n=30$ basis elements starting with an MPS of bond dimension $b_1$ 
(shown in legend) as a function of bond dimension $b_2$ (x-axis). }
\label{fig:approx_lanczos1}
\end{figure*}

\begin{figure}
\includegraphics[width=\linewidth]{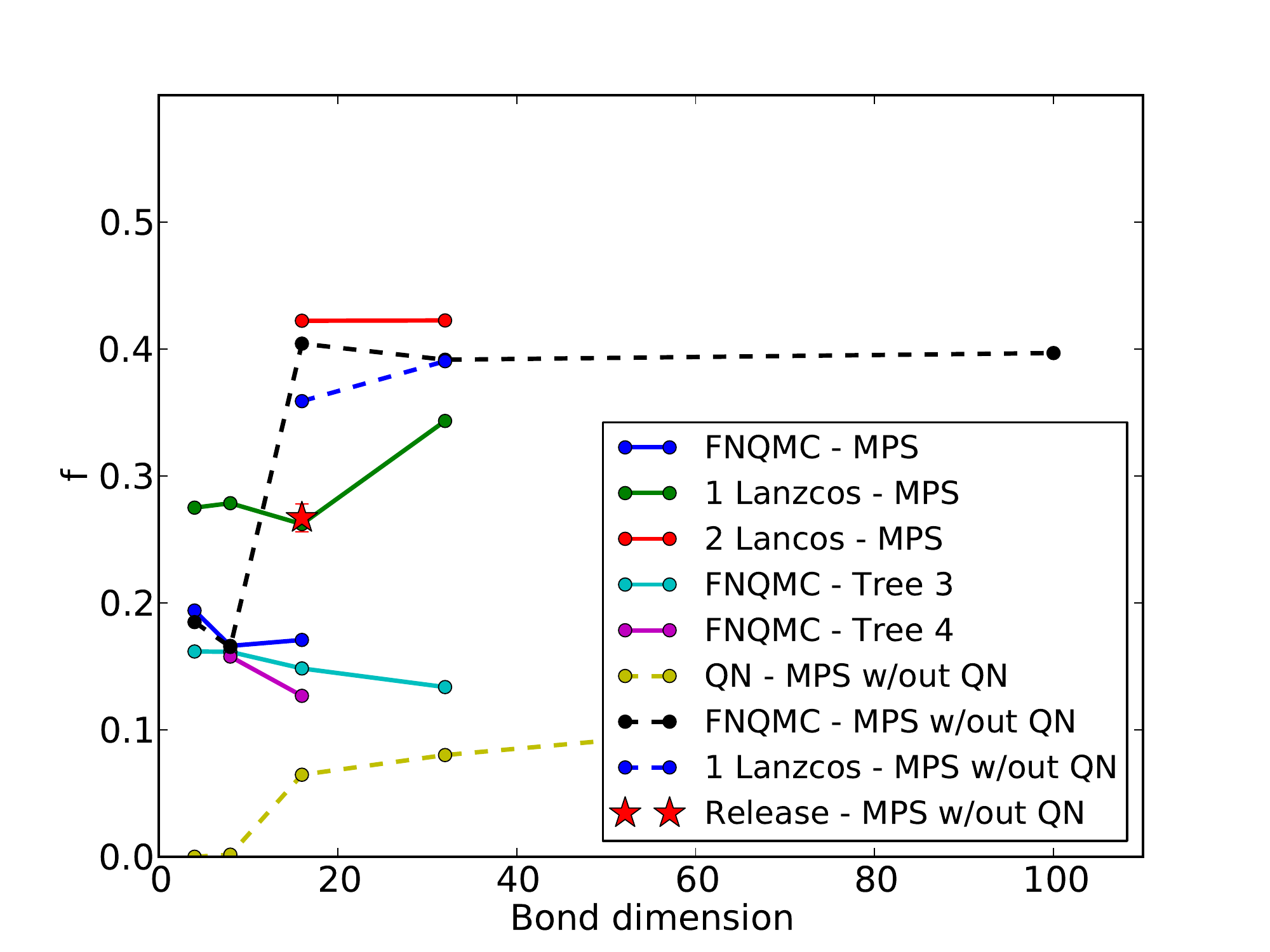}
\caption{Fraction of the energy $f$ (see eqn.~\eqref{eq:fraction_recovered}) recovered using 
various projection approaches on MPS and TTN as a function of bond dimension $D$.}
\label{fig:relative}
\end{figure}

One can alternately compute these terms using QMC.  With variational Monte Carlo, one can compute 
$\langle \Psi_T H^n \Psi_T\rangle$ by sampling configurations $R$ with probability 
$|\langle R | \Psi_T \rangle|^2$  and computing $\langle H^n |\Psi_T\rangle / | \Psi_T \rangle  \rangle_{|\Psi_T|^2}$.
The asymptotic scaling of these operations is $(6N)^{n}$ times the cost of contracting the tensor network
where $N$ is the number of sites. Hybrid approaches using some QMC and some MPO/MPS formalism are possible.
The Lanczos numbers reported here were computed partially in the MPO-MPS formalism and partially with a 
hybrid MPO-QMC approach.

As Figure~\ref{fig:relative} shows (with use of quantum numbers), 
the 1-Lanczos step results are better than the fixed node results 
(albeit more expensive), recovering approximately 30$\%$ of the missing energy. 
The 2-Lanczos does even better recovering $40 \%$. Interestingly, without 
quantum numbers, the Lanczos step does approximately as well as 
fixed-node but this underestimates the quality because our present implementation 
of this method does not perform a post projection onto the definite 
$S_z$ sector.

Finally, we note that because of the computational cost, it is difficult to go beyond a few Lanczos 
steps. This is essentially because we are finding optimal parameters in the exact basis of $H^n|\Psi\rangle$.
A possible alternative is to generate a basis where each basis element is a cheap approximation 
to $H^n|\Psi\rangle$.  One way to accomplish this is to repeatedly apply the MPO operator $H$ to $| \Psi\rangle$ of 
bond dimension $b_1$
truncating down to a fixed bond-dimension $b_2$ at each step using
the zip-up method \cite{stoudenmire2010minimally} (see fig.~\ref{fig:approx_lanczos1} (center and right)). The Hamiltonian and overlap matrices can then
be generated in this basis and a variational upper bound to the ground state can be found by solving a 
generalized eigenvalue problem.   This can do no better than DMRG on a  starting bond dimension
of $n b_2$.  Nonetheless, $n$ can in principle be quite large since one simply needs to compute the $n^2$ 
matrix elements.  Empirically, though, we find that using this approach, we
do not gain significant energies for $n>4$.  This is confirmed by looking at the (numerical)
rank of the system. 
Interestingly enough, we find that, at least until $b_1>16$, the value of $b_1$ doesn't
matter significantly.
We find a better energy using approximate Lanczos for $b_1=4$ and $b_2=64$ compared
to exact 1-Lanczos on $b_1=4$ which uses a MPS $H|\Psi\rangle$ of bond dimension
140.  To compete with the 2-Lanczos on $b_1=4$ which uses a MPS of bond dimension 4900, we need only
 $b_2=128$. At  $b_1=4$,$b_2=256$ we actually restore $57\%$ of the missing energy in the tensor network.

\section{Conclusion}
In conclusion, we have shown how to improve upon tensor networks 
using projection techniques in a way that is efficient and 
massively parallelizable (the QMC methods are all largely embarrassingly
parallel). We have tested 
stochastic exact projection, Fixed Node QMC, and 
Lanczos steps done in different ways using tensor networks. 
We summarize our findings  using the $10\times10$ triangular lattice (with cylindrical boundaries) 
nearest neighbor Heisenberg model as a test bed.
 
We have found that it is feasible to perform release node calculations 
starting with MPS of reasonably small bond dimensions. We 
released the walk after starting with a VMC calculation and 
gained about 20$\%$ of the missing energy, 
before the numerical sign problem took over. 
We expect to achieve further improvements 
by chaining projectors.  For example, one can start
 with a distribution drawn from a fixed-node or Lanczos calculation.

Next, we applied a sign problem free fixed-node QMC projector (for lattice systems) 
which is formulated to provide a variational upper bound to the energy. 
For MPS wavefunctions, this was found to be equivalent to effectively increasing $D$ by a factor of 
2 to 3. In order to provide a better starting wavefunction for low bond dimensions, 
we also used tree tensor networks of coordination 3 and 4. We believe that the gains 
reported in this paper can further be improved with optimized mappings of a 2D system 
onto a tree~\cite{Nakatani_Chan_TTN}. From a qualitative viewpoint, 
our results indicate that applying the fixed-node method to tensor networks 
such as PEPS (which have the correct entanglement laws and respect 
the symmetries of the lattice, but which scale unfavorably with $D$) should help in 
increasing the effective bond dimension. 
We also note the gains we achieved by post-projecting the quantum number free 
MPS onto the correct $S_z$ sector, 
which is carried out in QMC in a straightforward way. 

A third direction explored in this paper is the 1 (and 2) Lanczos step methods, 
which work by finding an optimal wavefunction in the basis spanned by $H \Psi$ (and $H^{2}\Psi$).
Higher powers of $H^n \Psi$ were also calculated approximately 
in a MPO framework. 
While the 1-Lanczos was found to be close to the fixed-node improvement, the 2-Lanczos appears to be generically 
better. In addition, we find significant improvement using the approximate Lanczos technique.

\emph{Added note:} During the completion of this work, 
a paper by Wouters et al.~\cite{Wouters} was posted to the arxiv exploring 
yet a fourth method
to combine QMC with projection, in this case the AFQMC method.
They find similar levels of improvements on a different model
and slightly smaller system sizes.

\section{Acknowledgement}
This research is part of the Blue Waters sustained-petascale computing project, 
which is supported by the National Science Foundation (award number ACI 1238993) 
and the state of Illinois. 
Blue Waters is a joint effort of the University of Illinois at Urbana-Champaign and 
its National Center for Supercomputing Applications. 
Computation was also done on Taub (UIUC NCSA).
We acknowledge support from grant DOE, SciDAC FG02-12ER46875.
We thank Cyrus Umrigar for discussions and critically reading the manuscript.
We thank Michael Kolodrubetz and Katie Hyatt for useful conversations and 
Miles Stoudenmire for help with ITensor. 

\bibliographystyle{prsty}
\bibliography{DMC_MPS}

\end{document}